\documentclass[twocolumn,           % Format : preprint, twocolumn
               showpacs,            % Pacs : showpacs, noshowpacs
               showkeywords,
               nopreprintnumbers,     % Preprint: preprintnumbers,
               aps,                 % Society: ...
               prd,          	    % Journal Style : pra, prb, prc, prd, pre,
               letterpaper,             % Size : a4paper, ...
               groupeaddress,      % Affiliation (Title) : groupedaddress,
               nofootinbib,         % Footnote: footinbib, nofootinbib
               tightenlines,        % Remove additional spaces in a line
               floats,floatfix      % Floating pictures and tables
               ]{revtex4-1}

\usepackage{graphicx}% Include figure files
\usepackage{dcolumn}% Align table columns on decimal point
\usepackage{bm}% bold math
\usepackage{amsmath}
\usepackage{amsfonts,amssymb}
\usepackage{color}
\usepackage{enumerate}

\begin{document}

\title{Gravitational collapse in brane-worlds revisited}

\author{Miguel A. Garc\'{\i}a-Aspeitia$^{1,2}$}
\author{C. Ortiz$^{2}$}
\author{J. C. L\'opez-Dominguez$^{2}$}
\author{Sinhue Hinojosa-Ruiz$^{2}$}
\email{aspeitia@fisica.uaz.edu.mx}
\affiliation{$^{1}$Consejo Nacional de Ciencia y Tecnolog\'ia, Av, Insurgentes Sur 1582. Colonia Cr\'edito Constructor, Del. Benito Ju\'arez C.P. 03940, M\'exico D.F. M\'exico}
\affiliation{$^{2}$Unidad Acad\'emica de F\'isica, Universidad Aut\'onoma de Zacatecas, Calzada Solidaridad esquina con Paseo a la Bufa S/N C.P. 98060, Zacatecas, M\'exico.}

\author{Mayra J. Reyes-Ibarra$^{3}$} 
%\email{mreyes@fis.cinvestav.mx}
\affiliation{$^{3}$Departamento de F\'isica, Centro de Investigaci\'on y de Estudios Avanzados del IPN. 2508,
San Pedro Zacatenco, 07360, Gustavo A. Madero, Ciudad de M\'exico, M\'exico}

\begin{abstract}
This paper is dedicated to revisit the modifications caused by branes in the collapse of a stellar structure under the Snyder-Oppenheimer scheme. Due to the homogeneity and isotropy of the model, we choose study the case of a closed geometry described by $k=1$, through the tool of dynamical systems. We revisit the different components of the star and its evolution during the stellar collapse, paying particular attention to the non-local effects and the quadratic terms of the energy momentum tensor that come from branes corrections. In the same vein we realize a phase portrait together with a stability analysis with the aim of obtain information about the attractors or saddle points of the dynamical system under different initial conditions in the density parameters, remarking the parameters that come from branes contributions.
\end{abstract}

\draft
\date{\today}
\maketitle

%%%%%%%%%%%%%%%%%%%%%%%%%%%%%%%%%%%%%%%%
\section{Introduction} \label{Int}
%%%%%%%%%%%%%%%%%%%%%%%%%%%%%%%%%%%%%%%%%

Brane-world theory, is one of the most promising extensions to General Relativity (GR) due of their natural ability to solve fundamental problems in physics with unprecedented success\cite{Maartens,PerezLorenzana:2004na,Anupam}. One of the problems addressed in this context, is for example, the hierarchy problem which is solved by the insertion of one or two $4$-dimensional branes in a $5$-dimensional bulk with Anti d'Sitter (AdS) geometry (see for example\cite{Gogberashvili:1998vx,Randall-I,Randall-II}) or a most complex bulk geometry\cite{Shiromizu}; due to this particular topology, the gravity is weakened as it approaches to our $4$-dimensional manifold (brane), solving in a natural way the problem of hierarchy\cite{Randall-I,Randall-II}. It is important to remark that this new topology generates modifications in Einstein's equations with the presence of second order term in the energy-momentum tensor, the presence of bulk matter and the existence of nonlocal terms due to the $5$-dimensional Weyl tensor\cite{Shiromizu}.

From here, it is noteworthy to remark the contributions to cosmology\cite{Binetruy,Wang,Binetruy:1999ut,aspeitia1,aspeitia2} and astrophysics\cite{Germani,Aspeitia3,Linares,Bruni:2001fd,Gergely:2011df,Garcia-Aspeitia:2014pna,Linares:2015fsa,Ovalle:2014uwa}, establishing new dynamics and alternative solutions to for example, the stellar stability, dark energy and dark matter problems\cite{Lue,PhysRevLett.90.241301,Okada:2004nc,Schwindt:2005fm,Chakraborty:2007ad,Sano:2009sp,Shahidi:2010nj}. In cosmological context, it is possible to find quadratic terms and non local terms, which are relevant at high energies, providing new dynamic in inflationary models\cite{Maartens:2000az,*Maartens:1999hf,*Gonzalez:2008wa} among others cosmological studies. 
Also we have new astrophysical dynamics from the brane-world point of view, where it is possible to explore the stability bounds generated by the presence of extra dimensions\cite{Germani,Garcia-Aspeitia:2014pna,Linares:2015fsa}, or the behavior of a star with a polytropic Equation of State (EoS)\cite{Castro:2014xza}. 

There have been advances in this approach, where it is studied the stellar stability imposing different exterior conditions\cite{Germani}, there are also studies where it is considered a star with constant energy density in which is demonstrated that Schwarzschild exterior is the extreme case possible(see for example\cite{Garcia-Aspeitia:2014pna}), also it is demonstrated the minimal setup to obtain a stellar configuration with the initial conditions of any EoS and with a Schwarzschild exterior\cite{Garcia-Aspeitia:2014pna}.
As a counterpart, it is possible to study stellar collapse using the Snyder-Oppenheimer (SO) model through a no-go theorem in order to explore whether it is possible to maintain a Schwarzschild exterior type\cite{Bruni:2001fd}. In the same vein, other studies focus in the gravitational collapse and black hole formation and evolution\cite{Casadio:2004nz} as well as the curvature corrections in the stellar collapse\cite{Kofinas:2004pt}.

Before we start, let us mention here some experimental constraints on braneworld models, most of them about the so-called brane tension, $\lambda$, which appears explicitly as a free parameter in the corrections of the gravitational equations mentioned above. As a first example we have the measurements on the deviations from Newton's law of the gravitational interaction at small distances. It is reported that no deviation is observed for distances $l \gtrsim  0.1 \, {\rm mm}$, which then implies a lower limit on the brane tension in the Randall-Sundrum II model (RSII): $\lambda> 1 \, {\rm TeV}^{4}$\cite{Kapner:2006si,*Alexeyev:2015gja}; it is important to  mention that these limits do not apply to the two-branes case of the  Randall-Sundrum I model (RSI) (see\cite{Maartens} for details). 
Astrophysical studies, related to gravitational waves and stellar stability, constrain
the brane tension to be  $\lambda > 5\times10^{8} \, {\rm MeV}^{4}$\cite{Germani,Sakstein:2014nfa}, whereas the existence of black hole X-ray binaries suggests that $l\lesssim 10^{-2} {\rm mm}$\cite{Maartens,Kudoh:2003xz,*Cavaglia:2002si}. Finally, from cosmological observations, the requirement of successful nucleosynthesis provides the lower limit $\lambda> 1\, {\rm MeV}^{4}$, which is a much weaker limit as compared to other experiments (another cosmological tests can be seen in: Ref.\cite{Holanda:2013doa,*Barrow:2001pi,*Brax:2003fv}).

Based in this background, this paper is dedicated to study the SO collapse in brane-world point of view through the theory of dynamical systems. We follow the standard procedure\cite{weinberg:1972} and it is extended to the case of branes with the aim of delve the behavior of the various components of the star at different stages of collapse, showing the critical points, stability and saddle points, remarking the advantage of this method to describe the dynamic of the star in its final stages\cite{Escobar:2011cz,Escobar:2012cq}.

This paper is organized as follows: in Sec. \ref{MF} we show the mathematical formalism necessary for the study of brane theory. In Sec. \ref{SO} we explore the model of SO collapse in brane context, performance the results through a numerical analysis. Finally in Sec. \ref{CR} we discuss our results obtained throughout the paper. Henceforth we use units in which $c=\hbar=1$.

%%%%%%%%%%%%%%%%%%%%%%%%%%%%%%%%%%%%%%%%
\section{Mathematical Background} \label{MF}
%%%%%%%%%%%%%%%%%%%%%%%%%%%%%%%%%%%%%%%%

Let us start by writing the equations of motion for stellar stability in a brane embedded in a 5D bulk according to the RSII model\cite{Randall-II}. Following an appropriate computation (for details see\cite{Maartens,*Shiromizu}), it is possible to demonstrate that the modified 4D Einstein's equation can be written as 
\begin{equation}
  G_{\mu\nu} + \Lambda_{(4)}g_{\mu\nu} = \kappa^{2}_{(4)} T_{\mu\nu} + \kappa^{4}_{(5)} \Pi_{\mu\nu} +
  \kappa^{2}_{(5)} F_{\mu\nu} - \xi_{\mu\nu}, \label{Eins}
\end{equation}
here $T_{\mu \nu}$ is the four-dimensional energy-momentum tensor of the matter in the brane, $\Lambda_{(4)}$ is the four dimensional cosmological constant and $\kappa_{(4)}$ is the four dimensional coupling constant which is related with the five dimensional coupling constant $\kappa_{(5)}$, through the relation $\kappa^{2}_{(4)}=8\pi G_{N}=\kappa^{4}_{(5)}\lambda/6$, where $\lambda$ is the brane tension parameter and $G_{N}$ is the Newton constant.
Also we have that $\Pi_{\mu\nu}$ represents the quadratic corrections on the brane generated from the four-dimensional energy-momentum tensor $T_{\mu \nu}$, whereas $F_{\mu\nu}$ gives the contributions of the energy-momentum tensor in the bulk $T_{AB}$ (with latin letters taking values $0,1,2,3,4$), which is then projected onto the brane with the help of the unit normal vector $n_{A}$. Finally, $\xi_{\mu\nu}$ gives the contributions of the five-dimensional Weyl's tensor $^{(5)}C^{E}_{AFB}$ when projected onto the brane manifold (see\cite{Shiromizu} for more details).

For simplicity, we will not consider bulk matter and then $T_{AB} = 0$, which translates into $F_{\mu\nu}=0$, and will also discard the presence of the four-dimensional cosmological constant, $\Lambda_{(4)} = 0$, as we do not expect it to have any important effect at astrophysical scales (for a recent discussion about this see\cite{Pavlidou:2013zha}). The energy-momentum tensor $T_{\mu \nu}$, the quadratic energy-momentum tensor $\Pi_{\mu \nu}$, and the Weyl (traceless) contribution $\xi_{\mu\nu}$, have the explicit forms:
\begin{subequations}
\begin{eqnarray}
\label{Tmunu}
T_{\mu\nu} = \rho u_{\mu}u_{\nu} + p h_{\mu\nu} \, , \\
\label{Pimunu}
\Pi_{\mu\nu} = \frac{1}{12} \rho [ \rho u_{\mu}u_{\nu} + (\rho+2p) h_{\mu\nu} ] \, , \\
\label{ximunu}
\xi_{\mu\nu} = -\left(\frac{\kappa_{(5)}}{ \kappa_{(4)}}\right)^{4} [\mathcal{U} u_{\mu}u_{\nu} + \mathcal{P}r_{\mu}r_{\nu} + \frac{h_{\mu\nu}}{3}(\mathcal{U}+\mathcal{P})] \, ,
\end{eqnarray}
\end{subequations}
where $p$ and $\rho$ are, respectively, the pressure and energy density of the stellar matter of interest, $\mathcal{U}$ is the nonlocal energy density, $\mathcal{P}$ is the nonlocal anisotropic stress, $u_{\alpha}$ the four-velocity (that also satisfies the condition $g_{\mu\nu}u^{\mu}u^{\nu}=-1$), $r_{\mu}$ is a unit radial vector and $h_{\mu\nu} = g_{\mu\nu} + u_{\mu} u_{\nu}$ is the projection operator orthogonal to $u_{\mu}$. In this case we are assuming spherical symmetry for a real star.

%%%%%%%%%%%%%%%%%%%%%%%%%%%%%%%%%%%%%%%%
\section{Snyder-Oppenheimer collapse in branes} \label{SO}
%%%%%%%%%%%%%%%%%%%%%%%%%%%%%%%%%%%%%%%%

We turn our attention to the gravitational collapse of a ball made of dust, and look for the dynamic caused by the brane corrections\cite{Bruni:2001fd}. Our arguments below will follow the simple assumptions we have made so far about the brane corrections on the $4$-dimensional gravitational equations of motion. To begin with, we find it convenient to write the physical basis of a comoving coordinate system in the form:
\begin{equation}
ds^{2}=-dt^{2}+U(r,t)dr^{2}+V(r,t)(d\theta^{2}+sin^{2}(\theta)d\varphi^{2}). \label{met}
\end{equation}
For practical porpoises, it is possible rewrite Eq. \eqref{Eins} in the form:
\begin{equation}
  R_{\mu\nu} = \kappa^2_{(4)} \left(T_{\mu\nu}-\frac{1}{2}g_{\mu\nu}T\right) +  \kappa^4_{(5)}\left(\Pi_{\mu\nu} - \frac{1}{2}g_{\mu\nu}\Pi\right) - \xi_{\mu\nu}, \label{Einsmod}
\end{equation}
then, the non-null terms are:
\begin{subequations}
\begin{eqnarray}
&&\frac{\dot{U}\dot{V}}{2UV}-\frac{\dot{U}^2}{4U^2}+\frac{\ddot{U}}{2U}-\frac{1}{U}\left[\frac{V^{\prime\prime}}{V}-\frac{V^{\prime2}}{2V^2}-\frac{U^\prime V^\prime}{2UV}\right]=4\pi G_N\rho_{eff}, \label{Rtt}\\
&&\frac{\dot{V}\dot{U}}{4VU}+\frac{\ddot{V}}{2V}+\frac{1}{V}-\frac{1}{U}\left[ \frac{V^{\prime\prime}}{2V}-\frac{U^{\prime}V^{\prime}}{4UV}\right]= 4\pi G_N\rho_{eff},\label{Rrr}\\
&&\frac{\dot{U}^2}{4U^2}+\frac{\dot{V}^2}{2V^2}-\frac{\ddot{U}}{2U}-\frac{\ddot{V}}{V} =4\pi G_N\rho_{eff}, \label{Rthth} \\
&&\frac{V^{\prime}\dot{V}}{2V^2}+\frac{\dot{U}V^{\prime}}{2UV}-\frac{\dot{V}^{\prime}}{V}=0, \label{Rtr}
\end{eqnarray}
\end{subequations}
where dots represents derivative with respect to $t$ and primes represents derivatives with respect to $r$. We have also that $\rho_{eff}$ and $p_{eff}$ are defined as:
\begin{eqnarray}
\rho_{eff}&=&\rho\left(1+\frac{\rho}{2\lambda}\right)+\frac{\mathcal{V}}{\lambda}, \\
p_{eff}&=&p\left(1+\frac{\rho}{\lambda}\right)+\frac{\rho^2}{2\lambda}+\frac{\mathcal{V}}{3\lambda}+\frac{\mathcal{N}}{\lambda},
\end{eqnarray}
being $\mathcal{V}=6\mathcal{U}/\kappa^{4}_{(4)}$ and  $\mathcal{N}=4\mathcal{P}/\kappa^{4}_{(4)}$. In order to demonstrate the general conditions for stellar collapse, we start considering the following separable solution: $U=R^{2}(t)f(r)$ and $V=S^{2}(t)g(r)$; notice that \eqref{Rtr} requires: $S(t)=R(t)$, where $f$ and $g$ can be normalized.

Under the argument that we are still free to redefine the radial coordinate as an arbitrary function of $r$ (see\cite{weinberg:1972} for details), it is possible to write: $U=R^{2}f(r)$, $V=R^{2}(t)r^{2}$. Using Eqs. \eqref{Rrr} and \eqref{Rthth} we have:
\begin{equation}
\frac{f^{\prime}}{rf^2}=-\frac{1}{r^2}+\frac{1}{fr^2}-\frac{f^\prime}{2rf^2}=-2k, \label{k10}
\end{equation}
where prime denotes derivative with respect to $r$. Solving Eq. \eqref{k10} we have $f(r)=(1-kr^2)^{-1}$, obtaining a spatially homogeneous and isotropic metric
\begin{equation} \label{metFLRW}
ds^{2}=-dt^2+R^2(t)\left[\frac{dr^2}{1-kr^2}+r^2(d\theta^2+\sin^2(\theta)d\varphi^2)\right],
\end{equation}
here $R(t)$ is the evolving scale factor of the star and $k$ is associated with the geometry of the stellar configuration in equivalence with cosmology. The 4-dimensional Bianchi identities implies $\nabla^{\nu} T_{\mu\nu}^{eff}=0$ and from the conservation equations we get $\nabla^{\nu} T_{\mu\nu}=0$, even more for a spatially homogeneous and isotropic Friedman model we have: $\nabla_{\nu}\mathcal{U}=\mathcal{P}_{\mu\nu}=0$ (see\cite{Maartens:2000fg} for details). Then, it is possible to extract information about $\rho(t)$ and $\mathcal{V}(t)$ through the following conservation equations: 
\begin{eqnarray}
\dot{\rho}+3\mathcal{H}\rho=0, \;\;\; \dot{\mathcal{V}}+4\mathcal{H}\mathcal{V}=0, \label{12}
\end{eqnarray}
here the dot represents derivative with respect to $t$ and we define $\mathcal{H}\equiv\dot{R}/R$. From Eqs. \eqref{12} we obtain $\rho(t)=\rho_{0}R(t)^{-3}$ and $\mathcal{V}(t)=\mathcal{V}_{0} R(t)^{-4}$. Substituting in \eqref{Rtt} and solving for $\dot{R}$ we have
\begin{equation}
\mathcal{H}^2=-k+\frac{\kappa^2_{(4)}}{3}\left[\frac{\rho_0}{R^3}\left(1+\frac{\rho_{0}}{2\lambda R^3}\right)+\frac{\mathcal{V}_{0}}{\lambda R^4}\right]. \label{R}
\end{equation}
Notice that in the GR regime ($\rho_{0}/\lambda\to0$, $\mathcal{V}_{0}/\lambda\to0$) with $k\neq0$, we recover the parametric solution of cycloid reported in the literature\cite{weinberg:1972}. In general, Eq. \eqref{R} can be solved by quadratures giving the following expression:
\begin{widetext}
\begin{eqnarray}
t-t_{0} &=& \frac{2 (-1)^{1/6} \alpha^{1/3}}{9 (3)^{1/4}\sqrt{\alpha+x^3}} 
\left[(-1)^{5/6}\left(\frac{(-1)^{1/3} x}{\alpha^{1/3}}-1\right)\right]^{1/2}\left[\frac{(-1)^{2/3}x^2}{\alpha^{2/3}}+\frac{(-1)^{1/3}x}{\alpha^{1/3}}+1\right]^{1/2}\times \nonumber\\&& F\left(\arcsin\left((3)^{-1/4}\sqrt{\left(-\frac{(-1)^{5/6}x}{\alpha^{1/3}}-(-1)^{5/6}\right)}\right),(-1)^{1/3}\right), \label{FIN}
\end{eqnarray}
\end{widetext}
where $\alpha\equiv-ax^6 + bx^2 + c$, being $a\equiv3/\kappa^2_{(4)}\rho_0$, $b\equiv\mathcal{V}_{0}/\rho_{0}\lambda$, $c\equiv\rho_0/2\lambda$, $x\equiv(3/\kappa^{2}_{(4)}\rho_0)^{1/2}R^3$ and $F(z,y)$ is the Elliptic function, under the assumption of a closed geometry $k=1$, in concordance with conventional wisdom.

%%%%%%%%%%%%%%%%%%%%%%%%%%%%%%%%%%%%%%%%
\subsection{Dynamical Analysis}
%%%%%%%%%%%%%%%%%%%%%%%%%%%%%%%%%%%%%%%%

To complement this study, we analyze the equation of motion from the metric element \eqref{metFLRW}, where the stellar components evolve during the collapse:
\begin{equation}
\mathcal{H}(t)^{2}=\rho_{eff}(t)-k\rho_{k}(t), \label{sist}
\end{equation}
where $\rho_{k}(t)=1/R(t)^{2}$. In addition, we propose the following dimensionless variables
\begin{eqnarray}
\Omega_{m}\equiv\frac{\bar{\rho}}{\mathcal{H}^{2}}, \; \Omega_{\mathcal{V}}\equiv\frac{\mathcal{V}}{\mathcal{H}^{2}\lambda}, \, 
\Omega_{\lambda}\equiv\frac{\rho}{2\lambda}, \;  \Omega_{k} \equiv\frac{\rho_{k}}{\mathcal{H}^{2}}  \,  , \label{eq:17b}
\end{eqnarray}
subject to the Friedman condition
\begin{equation}
1+k\Omega_{k}=\Omega_{m}(1+\Omega_{\lambda})+\Omega_{\mathcal{V}}. \label{FC}
\end{equation}
The dynamical system can be written as:
\begin{subequations}
\begin{eqnarray}
\frac{\Omega_{m}^{\prime}}{\Omega_{m}}&=&-3-2\Pi, \;\; \frac{\Omega_{\mathcal{V}}^{\prime}}{\Omega_{\mathcal{V}}}=-4-2\Pi, \label{diffeq1}\\
\frac{\Omega_{\lambda}^{\prime}}{\Omega_{\lambda}}&=&-3, \;\; \frac{\Omega_{k}^{\prime}}{\Omega_{k}}=-2-2\Pi,
\label{diffeq}
\end{eqnarray}
\end{subequations}
where now, the primes denote derivative with respect to the $e$-foldings ($N=\ln(R)$) and $\Pi$ is defined as
\begin{equation}
\Pi\equiv\frac{\mathcal{H}^{\prime}}{\mathcal{H}}=-\frac{3}{2}\Omega_{m}(1+2\Omega_{\lambda})-2\Omega_{\mathcal{V}}+k\Omega_{k}, \label{H}
\end{equation}
which can be obtained from Eq. \eqref{sist}. From conditions \eqref{eq:17b} and \eqref{FC}, it is enough to investigate the flow of \eqref{diffeq1}-\eqref{diffeq} defined in the space phase
\begin{eqnarray}
\Psi&=&\lbrace(\Omega_{m},\Omega_{\lambda},\Omega_{k}): 0\leqslant\Omega_{m}(1+\Omega_{\lambda})-k\Omega_{k}\leqslant1,\nonumber\\&&0\leqslant\Omega_{m}\leqslant1, 0\leqslant\Omega_{\lambda}\leqslant1, 0\leqslant\Omega_{k}\leqslant1\rbrace,
\end{eqnarray}
where it is reduced the above condition due that we assume $0\leqslant\Omega_{\mathcal{V}}\leqslant1$.

\begin{figure}[htbp]
\centering
\begin{tabular}{cc}
\includegraphics[scale=0.6]{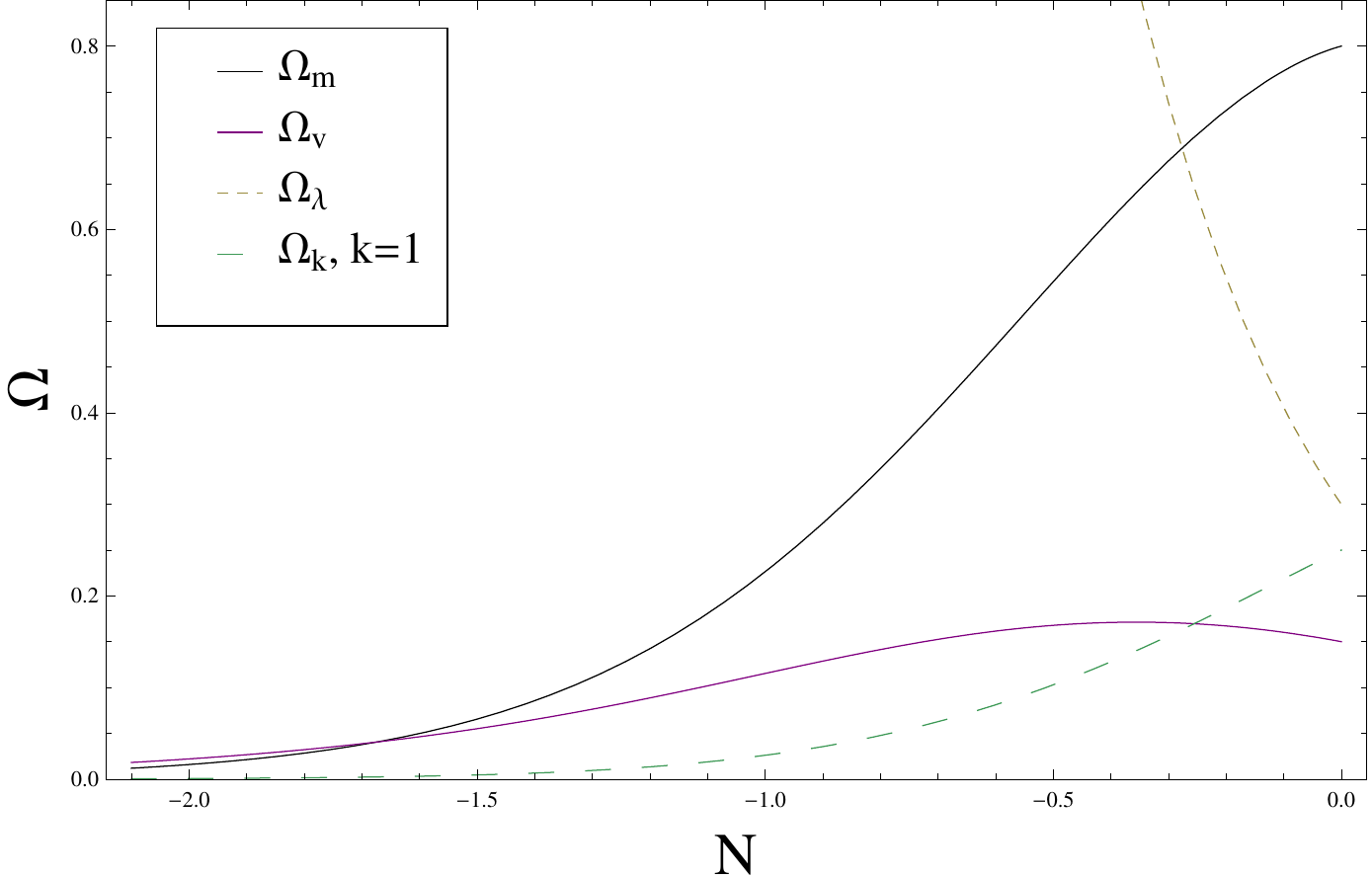} \\
\includegraphics[scale=0.6]{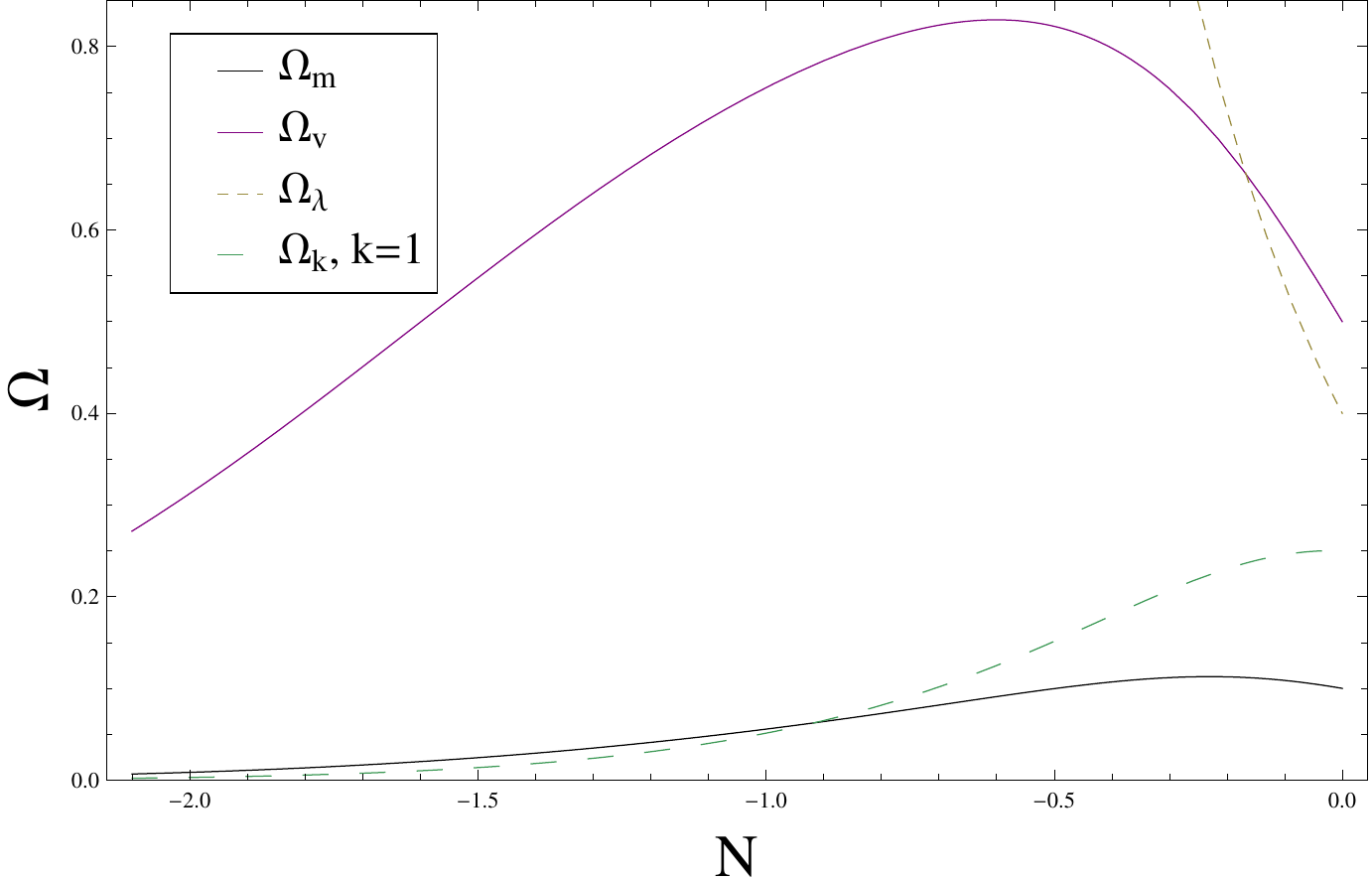}
\end{tabular}
\caption{Phase space of the system \eqref{diffeq1}-\eqref{diffeq}, for the spherical geometry $k=1$. (Top) Here we assume dominance of baryonic density parameter in comparison with the other components, as expected in a real stellar configuration. It is shown the matter, Weyl, brane tension and curvature terms during the evolution of collapse of the star. Notice how the high energy term $\Omega_{\lambda}$ always dominate in the closest moments of collapse. (Bottom) In this case we assume that brane corrections are dominant over the other components, maintaining the presence of non local terms until the final stages of collapse; the brane tension parameter always grow exponentially as we expect.}
\label{fig1}
\end{figure}

Before to start, we establish two important limits: The first one, is the \emph{low energy limit} where GR is recovered; in this case we have that $\Omega_{\lambda}\ll1$ and $\Omega_{\mathcal{V}}\sim0$ recovering the SO collapse obtained form classical GR. In the other case, brane effects begin to be significant, dominating over the other terms. Then, Eq. \eqref{H} and the Friedman constraint can be expressed as: $1+k\Omega_{k}=\Omega_{m}\Omega_{\lambda}+\Omega_{\mathcal{V}}$ together with 
\begin{equation}
\Pi_{High}=-3\Omega_{m}\Omega_{\lambda}-2\Omega_{\mathcal{V}}+k\Omega_{k}. \label{HE}
\end{equation}
Returning to our analysis, Eqs. \eqref{diffeq1}-\eqref{H} are readily soluble analytically, under the assumption $k=0$ due to the spherical geometry of the stellar configuration. The results are shown in Fig. \ref{fig1}. In the first case (Fig. \ref{fig1} Top) we have that the initial conditions imposed are: $\Omega_{m0}=0.8$, $\Omega_{\mathcal{V}0}=0.15$, $\Omega_{\lambda0}=0.3$ and $\Omega_{k0}=0.25$, where it is possible to observe the domination of baryonic matter. In the second case (Fig. \ref{fig1} Bottom) brane parameters are dominant over the other components, having the initial conditions: $\Omega_{m0}=0.1$, $\Omega_{\mathcal{V}0}=0.5$, $\Omega_{\lambda0}=0.4$ and $\Omega_{k0}=0.25$. It is possible to observe that the non-local term is always dominant until the final stages of the stellar collapse. Both initial conditions are chosen by hand, only by the premise of show the behavior of the components under different conditions.

It is important to notice, how Fig. \ref{fig1} show that the brane term $\Omega_{\lambda}$ grow exponentially in comparison with the other components; this behavior is due that the direct integration, generates an exponential function described in the following way: $\Omega_{\lambda}=\Omega_{\lambda0}\exp(-3N)$, notice how the other components are negligible nearest to the collapse.

\begin{figure}[htbp]
\centering
\begin{tabular}{cc}
\includegraphics[scale=0.83]{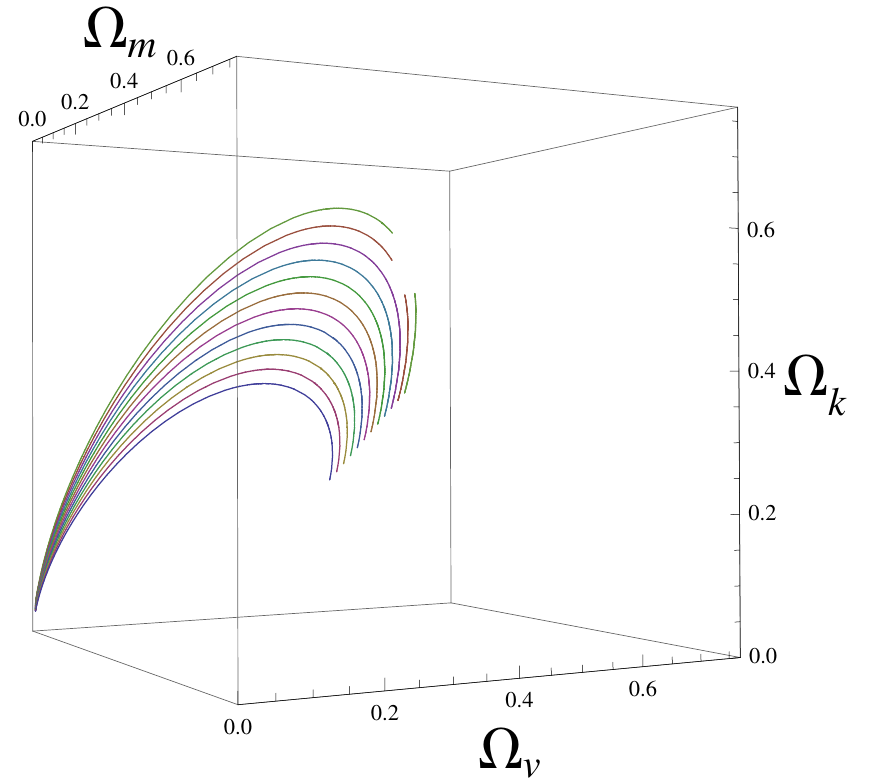} \\
\includegraphics[scale=0.7]{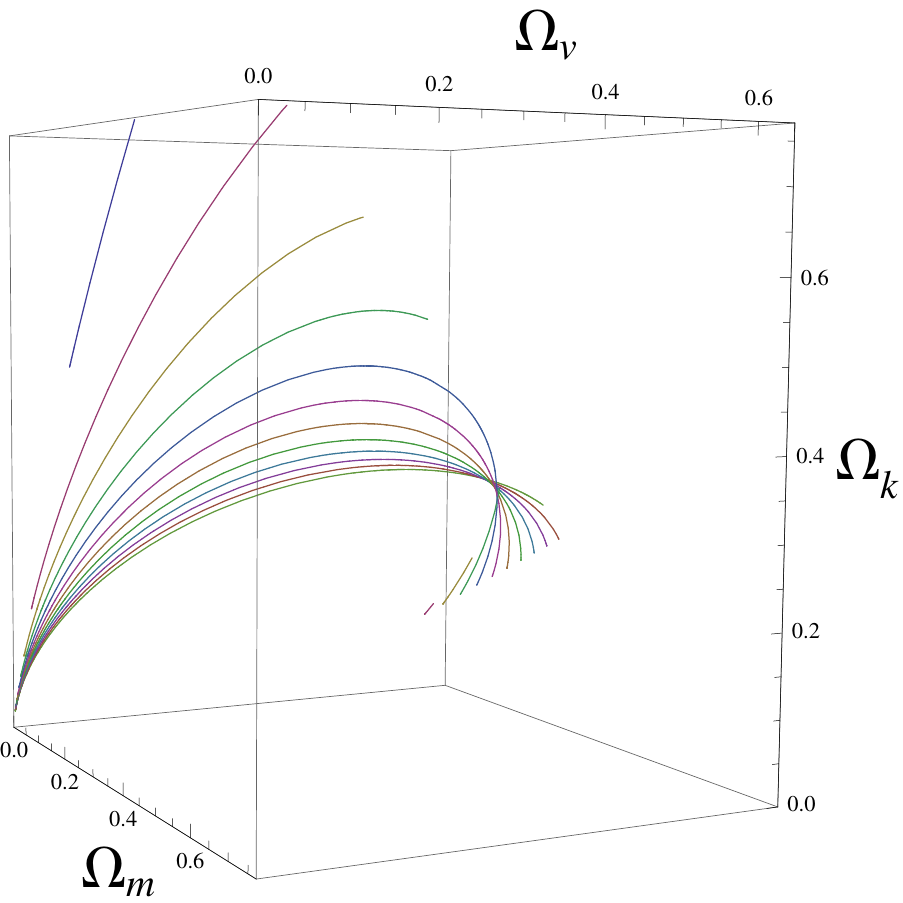}
\end{tabular}
\caption{Phase portrait of Eqs. \eqref{diffeq1}-\eqref{diffeq}. (Top) As initial conditions, we fix the value of the matter density parameter at $\Omega_{m}=0.7$ and the other two density parameters vary in the interval $\Omega_{k}\epsilon[0.2,0.4]$ and $\Omega_{\lambda}\epsilon[0.2,0.4]$ always maintaining the dominance of the matter density parameter. (Bottom) In this case we fix $\Omega_{k}=0.25$ and varying the other two parameters in the intervals $\Omega_{m}\epsilon[0.25,0.8]$ and $\Omega_{\lambda}\epsilon[0.15,0.7]$, from here it is possible to see two saddle points unlike the previous case.}
\label{fig2}
\end{figure}

To complement our study realized through this section, we analyze the equilibrium points and eigenvalues associated to Eqs. \eqref{diffeq1} and \eqref{diffeq} in order to obtain important information about the collapse behavior. It is possible to define the equation of critical points as: $(d\Omega_i/dN)_{\bf{x_0}} =0$. In this case we have three critical points with matter, Weyl fluid and curvature domination, shown in Table \ref{tab:critical}. Also, we define the vector $\textbf{x}=(\Omega_{m},\Omega_{\mathcal{V}},\Omega_{\lambda},\Omega_{k})$ and consider a linear perturbation of the form $\textbf{x}\to\textbf{x}_{c}+\delta\textbf{x}$. The linearized system reduces to $\delta\textbf{x}^{\prime}=\mathcal{J}\delta\textbf{x}$, where $\mathcal{J}$ is the Jacobian matrix of $\textbf{x}^{\prime}$, written as:
\begin{widetext}
\begin{equation}
\mathcal{J}=\left(\frac{\partial \Omega_{i}^{\prime}}{\partial \Omega_{j}}\right)_{\textbf{x}_{0}}=\begin{pmatrix}
-3-2[\Pi_{0}-\frac{3}{2}\Omega_{m0}(1+2\Omega_{\lambda0})] & 4\Omega_{m0} & 6\Omega_{m0}^{2} & -2k\Omega_{m0} \\
3\Omega_{\mathcal{V}0}(1+2\Omega_{\lambda0}) & -4-2(\Pi_{0}-2\Omega_{\mathcal{V}0}) & 6\Omega_{\mathcal{V}0}\Omega_{m0} & -2k\Omega_{\mathcal{V}0} \\
0 & 0 & -3 & 0 \\
3\Omega_{k0}(1+2\Omega_{\lambda0}) & 4\Omega_{k0} & 6\Omega_{k0}\Omega_{m0} & -2-2(\Pi_{0}+k\Omega_{k0}) \\
\label{Jacob}
\end{pmatrix}.
\end{equation}
\end{widetext}
In order to obtain information about attractors, saddle points and others, we found the eigenvalues associated with the Jacobian matrix \eqref{Jacob}. Table \ref{tab:critical} summarize the eigenvalues associated with this particular model and establish the conditions of the eigenvalues existence (see the middle part of Table \ref{tab:critical}). Our results remark the existence of saddle-like points in dynamics which represents unstable manifolds, being notorious in Figs. \ref{fig2} which are phase portrait of Eqs. \eqref{diffeq1}-\eqref{diffeq}. 

\begin{figure}[htbp]
\centering
\begin{tabular}{cc}
\includegraphics[scale=0.7]{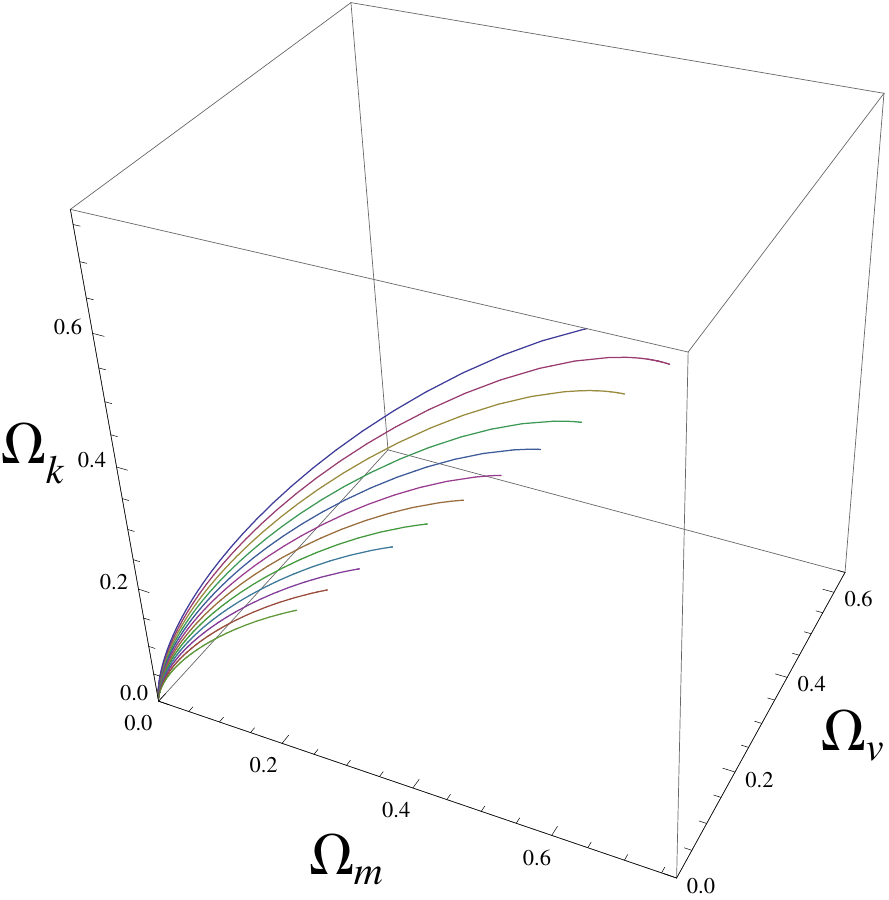}
\end{tabular}
\caption{Phase portrait of Eqs. \eqref{diffeq1}-\eqref{diffeq}. we apply the initial conditions $\Omega_{\mathcal{V}0}=0$ with $\Omega_{m0}\epsilon[0.25,0.8]$ and $\Omega_{k0}\epsilon[0.25,0.8]$, such that the only term in which it contributes branes is $\Omega_{\lambda0}=0.3$.}
\label{fig3}
\end{figure}

Figs. were computed under the following Initial conditions: In Fig. \ref{fig2} top we fix the value of the matter density parameter at $\Omega_{m0}=0.7$ and the other two density parameters vary in the interval $\Omega_{\mathcal{V}0} \epsilon [0.2,0.4]$ and $\Omega_{k0} \epsilon [0.2,0.4]$ always maintaining the dominance of the matter density parameter. In the same vein, for Fig. \ref{fig2} bottom, we analyze the behavior, fixing $\Omega_{k0}=0.25$ and varying the other two parameters in the intervals $\Omega_{m0} \epsilon [0.25,0.8]$ and $\Omega_{\mathcal{V}0} \epsilon [0.15,0.7]$. In both cases we fix the brane tension density parameter as: $\Omega_{\lambda0}=0.3$. Finally in Fig. \ref{fig3} we apply the initial conditions $\Omega_{\mathcal{V}0}=0$ with $\Omega_{m0}\epsilon[0.25,0.8]$ and $\Omega_{k0}\epsilon[0.25,0.8]$, such that the only term in which it contributes branes is $\Omega_{\lambda0}=0.3$.

\begin{table}[htp]
\caption{\label{tab:critical} Critical points for the system \eqref{diffeq1} and \eqref{diffeq}. From left to right, the columns read: Point, coordinates, existence, eigenvalues and stability.}
\begin{ruledtabular}
\begin{tabular}{c c c c c c}
Point & Coordinates & Existence & Eigenvalues & Stability \\
$P_1$ & $(1,0,0,0)$ & all $k$ & $(1,-3,3,-1)$ & Saddle point  \\
$P_2$ & $(0,1,0,0)$ & all $k$ & $(0,1,2,-3)$ & Saddle point  \\
$P_3$ & $(0,0,0,-1/k)$ & $k=\pm1$ & $(2,-3,-2,-1)$ & Saddle point 
\end{tabular}
\end{ruledtabular}
\end{table}

%%%%%%%%%%%%%%%%%%%%%%%%%%%%%%%%%%%%%%%%
\section{Discussion} \label{CR}
%%%%%%%%%%%%%%%%%%%%%%%%%%%%%%%%%%%%%%%%%

In this paper we implement a roboust analysis of the SO collapse in the background of branes showing the new terms that will play a role in the dynamic of collapse of a star. We start with the equation of motion for SO collapse which is shown in Eq. \eqref{R} and solved analytically for the most general case where branes play an important role, never losing sigh that Eq. \eqref{R} and Eq. \eqref{FIN} converge to GR in the appropriate limit $\rho_{0}/\lambda\to0$ and $\mathcal{V}_{0}/\lambda\to0$.

Similar analysis was conducted from the point of view of dynamical systems in order to study the dynamics of the different components of the star. In this case, we focused on the matter components (dust), the brane terms thats grows proportional to $\rho/\lambda$, the non-local terms due to the Weyl tensor and the geometrical term fixed only in the spherical geometry $k=1$. From this study, it was possible to extract relevant information regarding the behavior of stellar collapse with the addition of terms that come from brane-worlds. In this case, we impose reasonable conditions that must contain a star in the most natural possible conditions. As it was expected, the brane term associated with $\Omega_{\lambda}$ must dominate over the other density parameters due to its exponential behavior, allowing the natural decaying of the other components in the final stages of the stellar collapse. Notoriously, Weyl terms only dominates slightly in the final stage but always is subdominant previous the collapse as we expect from the traditional knowledge of stellar dynamics. On the other hand, we explore the dynamic when the brane components are dominant over the other components, showing the dominance of the Weyl term even in the final stages of the collapse. 

To complement, we realize the phase portrait of Eqs. \eqref{diffeq1}-\eqref{diffeq} (see Figs. \ref{fig2} and \ref{fig3}) assuming different initial conditions with the aim of prove the behavior under different conditions. The mathematical and numerical analysis show saddle points which are unstable in the dynamics, noticing that the presence of non-local terms generates the discontinuity presented in Figs. \ref{fig2}. From Fig. \ref{fig3} the plot is only restricted to the plane $\Omega_{k}$, $\Omega_{m}$ due to the condition $\Omega_{\mathcal{V}0}=0$; despite this, the presence of branes still there, mediated by the term $\Omega_{\lambda}$.

As a final note, we emphasize that the procedure can be replicated for other classes of stars that may include the presence of polytropic matter or with other most general focuses and whose solutions may also require extra numerical analysis. However this is ongoing research that will be presented elsewhere.

%%%%%%%%%%%%%%%%%%%%%%%%%%%%%%%%%%%%%%%%%%%%%%%
\begin{acknowledgements}
The authors acknowledge support from SNI-M\'exico, PIFI and PROMEP with number CA 205. MAG-A acknowledge support from CONACyT research fellow, Instituto Avanzado de Cosmolog\'ia (IAC) collaborations. JCL-D acknowledge support from F-PROMEP-39/Rev-03. MJR-I acknowledges support from a CONACyT postdoctoral grant.
\end{acknowledgements}

\bibliography{librero0}

\end{document}